\documentstyle[aps,epsfig]{revtex}  
\begin{document}    
\draft    
\title{Andreev reflection at QGP/CFL interface }   
\author{M. Sadzikowski$^{1,2}$, 
        M. Tachibana$^{1,3}$\footnote{JSPS Research Fellow} }   
\address{ 1) Center for Theoretical Physics, Massachusetts Institute of Technology,
             Cambridge MA 02139-4307, USA \\
          2) Institute of Nuclear Physics, Radzikowskiego 152, 
             31-342 Krak\'ow, Poland\\
          3) Yukawa Institute for Theoretical Physics, Kyoto University, Kyoto 606-8502, Japan}    
\maketitle    
\tighten    
\begin{abstract} 
In this letter we address the question of the phenomena
of Andreev reflection between the cold quark-gluon plasma
phase and CFL color superconductor. We show that
there are two different types of reflections connected
to the structure of the CFL phase. We also calculate
the probability current at the interface and we show
that it vanishes for energy of scattering quarks
below the superconducting gap.
\end{abstract}  
%%%%
\section{Introduction}
The prediction of the existence of the superconducting
phase as a true vacuum for high density QCD
\cite{2sc} raises the interesting possibility
that this phase can be found inside the Neutron Stars.
There are a lot of papers which try to find the
possible observable effects that can be 
caused by the superconducting phase (see for
review \cite{rw}). 

In this letter we try to point out on the phenomenon
that take place at the interface between the quark-gluon
plasma and CFL phase \cite{cfl} which can be of great importance
for transport phenomena of the matter inside
the Neutron Star. There is a well-known process
of Andreev reflection \cite{andreev} in condensed matter
systems that can also take place in the case
of color superconductors. This possibility was mentioned
in the paper \cite{mariusz1} for the case of two 
light flavor superconductivity (2SC phase).
Here we consider in detail the Andreev reflection
process for CFL supercoductors. 

In the second section
we describe the effective hamiltonian and later
we discuss the influence of Nambu-Goldstone modes
for the reflection process. The fourth section
contains the detail discussion of Andreev reflection
and formula for probability current.
%%%%%%%%%%%%%%%%%%%%%%%%%%%%%%%%%%%%%%%%%%%%%%%%%%%%%%%%%%%%%%%%%%%%%%%%%%%%%%%%%%%%%%%%%%%%%
\section{Effective Hamiltonian}
In QCD at asymptotically high density, the dominant interaction
between quarks is carried by one gluon exchange, in which the
interaction force is attractive in the color $\bar{3}$ channel.
The dominant coupling is a color and flavor anti-symmetric
interaction of the form from one gluon exchange\cite{schafer}
%%%%
\begin{equation}
{\cal L}_{eff} = G(\delta^{ac}\delta^{bd}-\delta^{ad}\delta^{bc})
(\delta_{ik}\delta_{jl}-\delta_{il}\delta_{jk})
(\psi^{a{\rm T}}_i C\gamma_5 \psi^{b}_j )(\psi^{c{\rm T}}_k C\gamma_5 \psi^{d}_l )^{\dagger}. 
\label{lag1}
\end{equation}
%%%%
where $a, b, c$ and $d$ are color indices and $i, j, k$ and $l$ are
flavor indices. $\psi^a_{i}$ is the quark field operator.

Now let us assume that the condensate takes the form
%%%%
\begin{equation}
4G<\psi^{a{\rm T}}_i C\gamma_5 \psi^{b}_j> = \Delta^{ab}_{ij}.
\label{condensate}
\end{equation}
%%%%
Here $\Delta^{ab}_{ij}$ is anti-symmetric under the exchanges
of $a \leftrightarrow b$ and $i \leftrightarrow j$, respectively.

Using eq.(\ref{condensate}) one obtains at the mean field level:
%%%%
\begin{equation}
{\cal L}_{eff} = 
\left(\Delta^{ab}_{ij}(\psi^{c{\rm T}}_k C\gamma_5 \psi^{d}_l )^{\dagger} + h.c.\right)
+ \tilde{{\cal L}}.
\label{lag2}
\end{equation}
%%%%
where $\tilde{{\cal L}}$ is independent of the fermionic fields.

The effective hamiltonian of the system of interest is
%%%%
\begin{equation}
{\rm H_{eff}} = \int d^3x \left[\psi^{a\dagger}_i
(-i\vec{\alpha}\cdot\vec{\nabla}+m\gamma_0-\mu)\psi^a_i
+\left(\Delta^{ab}_{ij}(\psi^{a{\rm T}}_i C\gamma_5 \psi^{b}_j )^{\dagger} + h.c.
\right)\right]
\label{effham}
\end{equation}
%%%%
with $\vec{\alpha} \equiv \gamma_0\vec{\gamma}$.
Here $m$ is the current quark mass and $\mu$ is the quark chemical potential.
$\Delta$ depends on the positions: it vanishes in the QGP phase and takes
non-zero value in CFL phase.

In CFL phase, color and flavor degrees of freedom are locked to
each other. So the gap matrix $\Delta^{ab}_{ij}$, which is a $3 \times 3$
matrix in both color and flavor spaces, takes the form \cite{cfl}
%%%%
\begin{equation}
\Delta^{ab}_{ij} = \Delta(\delta^{a}_{i}\delta^{b}_{j}-\delta^{a}_{j}\delta^{b}_{i}).
\label{gap}
\end{equation}
%%%%
Here we have neglected the contribution from color 6 channel, which is symmetric
in color and flavor since it was shown to be small \cite{cfl}.

From the structure of the gap matrix (\ref{gap}), we find that there are
two kinds of quark pair structures. One consists of only two quark contribution, such as
%%%%%%%%%
\begin{eqnarray}
\left(\begin{array}{c} u_{green}\\ d_{red}\\ 0\end{array}\right), \quad
\left(\begin{array}{c} u_{blue}\\ 0\\ s_{red}\end{array}\right) \quad {\rm and} \quad
\left(\begin{array}{c} 0\\ d_{blue}\\ s_{green}\end{array}\right). 
\label{2SC}
\end{eqnarray}
%%%%%%%%%%
These cases are similar to 2SC phase, where we have only two flavors. 
On the other hand in the CFL phase there is another structure of quark pairs composed of
%%%%%%%%%
\begin{eqnarray}
\left(\begin{array}{c} u_{red}\\ d_{green}\\ s_{blue}\end{array}\right),
\label{CFL}
\end{eqnarray}
%%%%%%%%
which never appears in 2SC phase \footnote{ This quark pair structures have
been discussed in \cite{unlock}.}. 
Let us call it the triplet
for short. We will mainly restrict our consideration of Andreev reflection 
into this case because the 2SC-like pairs have been already discussed in \cite{mariusz1}.
%%%%%%%%%%%%%%%%%%%%%%%%%%%%%%%%%%%%%%%%%%%%%%%%%%%%%%%%%%%%%%%%%%%%%%%%%%%%%%%%%%%%%%%%%%
\section{Nambu-Goldstone modes}
Before going into details, let us mention about another aspect of
CFL phase. Unlike 2SC phase, there exist Nambu-Goldstone (NG) bosons
associated with chiral and baryon symmetry breaking in CFL phase. 
Because they are the lowest order excitations one can think that
they can influence in large extent the Andreev reflection process.
However we find the NG bosons do not play an important role in the current problem.

The effective interaction in the lowest order between fermions and NG bosons
in CFL phase is given by the formula\cite{casal}:
%%%%%%%%
\begin{eqnarray}
{\rm H^{int}} &=& {\rm H}_{\chi\chi\Pi} + {\rm H}_{\chi\chi\Pi\Pi} \nonumber \\
      &=& \chi^{\dagger}\left(\begin{array}{cc}
0 & \Delta\left(\frac{2i}{F_{\Pi}}\Pi-\frac{2}{F^2_{\Pi}}\Pi^2\right)\\
\Delta^*\left(-\frac{2i}{F_{\Pi}}\Pi-\frac{2}{F^2_{\Pi}}\Pi^2\right) & 0
\end{array}\right)\chi,
\label{ng1}
\end{eqnarray}
%%%%
where $\chi$ is the Nambu-Gorkov fermion field and $\Pi$ is the Nambu-Goldstone
field. $F_{\Pi}$ is the decay constant of the NG field. 
From this form of the interaction we find the following Feynman rules (Fig.1):
%%%%%%%%
\par
\begin{center}
\leavevmode{\epsfxsize=10cm\epsffile{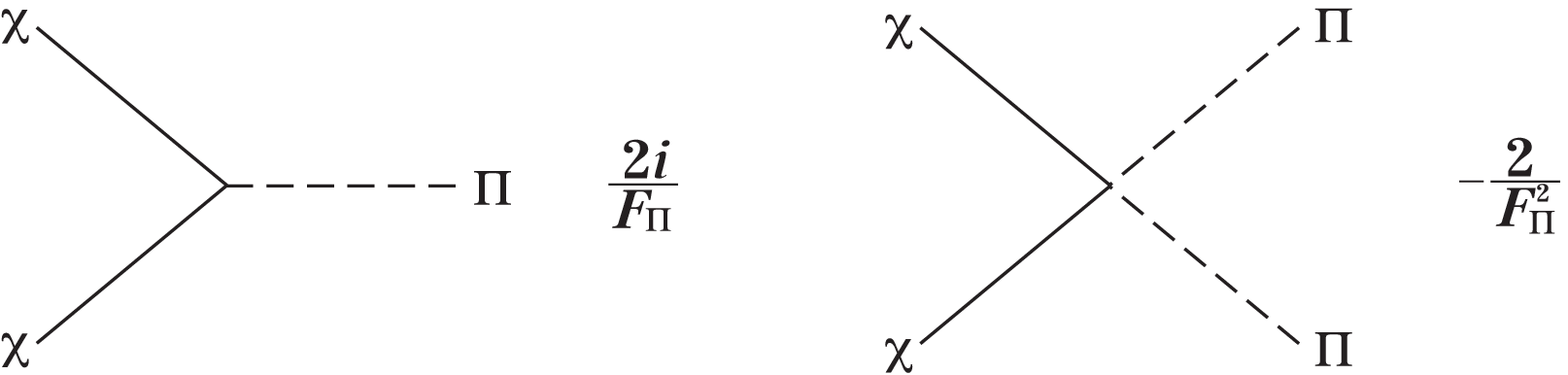}}\\
\par
Fig.1 The lowest order interaction vertices between quarks and the NG bosons.
\end{center}
\smallskip
\par
%%%%%%%
As has been obvious from these diagrams, the interaction between quarks
and the NG bosons is suppressed by the factor $\frac{1}{F_{\Pi}}$. On the
other hand, $F_{\Pi}$ is of the order of ${\cal O}(\mu)$\cite{son}. 
As the result, the emission of the  NG bosons from quarks is highly suppressed at high density. 
So we can neglect the effect of the NG bosons at leading order in $\mu$ expansion.
%%%%%%%%%%%%%%%%%%%%%%%%%%%%%%%%%%%%%%%%%%%%%%%%%%%%%%%%%%%%%%%%%%%%%%%%%%%%%%%%%%%%%%%%%%
\section{Andreev reflection}
Let us now discuss Andreev reflection at QGP/CFL interface for the
triplet state we mentioned in the chapter II. We choose $u_{red}$, $d_{green}$ and
$s_{blue}$ basis because it is natural in the QGP phase.
The equations of motions that follow from (\ref{effham}) take the form: 
%%%%%
\begin{eqnarray}
i\dot{\psi}^{u}_{red} &=& (-i\vec{\alpha}\cdot\vec{\nabla}+m\gamma_0-\mu)\psi^{u}_{red}
-\Delta C\gamma_5 (\psi^{d*}_{green}+\psi^{s*}_{blue}), \nonumber \\
i\dot{\psi}^{d}_{green} &=& (-i\vec{\alpha}\cdot\vec{\nabla}+m\gamma_0-\mu)\psi^{d}_{green}
-\Delta C\gamma_5 (\psi^{u*}_{red}+\psi^{s*}_{blue}), \nonumber \\
i\dot{\psi}^{s}_{blue} &=& (-i\vec{\alpha}\cdot\vec{\nabla}+m\gamma_0-\mu)\psi^{s}_{blue}
-\Delta C\gamma_5 (\psi^{u*}_{red}+\psi^{d*}_{green}), \nonumber \\
i\dot{\psi}^{u\dagger}_{red} 
&=& -i\vec{\nabla}\psi^{u\dagger}_{red}\cdot\vec{\alpha}-\psi^{u\dagger}_{red}(m\gamma_0-\mu)
-\Delta^* (\psi^{d\rm{T}}_{green}+\psi^{s\rm{T}}_{blue})C\gamma_5, \nonumber \\
i\dot{\psi}^{d\dagger}_{green} 
&=& -i\vec{\nabla}\psi^{d\dagger}_{green}\cdot\vec{\alpha}-\psi^{d\dagger}_{green}(m\gamma_0-\mu)
-\Delta^* (\psi^{u\rm{T}}_{red}+\psi^{s\rm{T}}_{blue})C\gamma_5, \nonumber \\
i\dot{\psi}^{s\dagger}_{blue} 
&=& -i\vec{\nabla}\psi^{s\dagger}_{blue}\cdot\vec{\alpha}-\psi^{s\dagger}_{blue}(m\gamma_0-\mu)
-\Delta^* (\psi^{u\rm{T}}_{red}+\psi^{d\rm{T}}_{green})C\gamma_5.
\label{eom}
\end{eqnarray}
%%%%%

To find the quasiparticle wavefunctions for $\Delta = const$, 
it is convenient to use the following decompositions:
%%%%%
\begin{eqnarray}
\psi^u_{red}(t, \vec{r}) = \sum_{r}\alpha_r\varphi^{u}_{r,R}(\vec{q})\exp(i\vec{q}\cdot\vec{r}-iEt),
& & \quad
\psi^{u\dagger}_{red}(t, \vec{r}) 
= \sum_{r}\alpha_r^* h^{u\dagger}_{r,L}(-\vec{q})\exp(i\vec{q}\cdot\vec{r}-iEt),
\nonumber \\
\psi^d_{green}(t, \vec{r}) = \sum_{r}\beta_r \varphi^{d}_{r,R}(\vec{q})\exp(i\vec{q}\cdot\vec{r}-iEt),
& & \quad 
\psi^{d\dagger}_{green}(t, \vec{r}) 
= \sum_{r}\beta_r^* h^{u\dagger}_{r,L}(-\vec{q})\exp(i\vec{q}\cdot\vec{r}-iEt),
\nonumber \\
\psi^s_{blue}(t, \vec{r}) = \sum_{r}\gamma_r \varphi^{s}_{r,R}(\vec{q})\exp(i\vec{q}\cdot\vec{r}-iEt),
& & \quad
\psi^{s\dagger}_{blue}(t, \vec{r}) 
= \sum_{r}\gamma_r^* h^{u\dagger}_{r,L}(-\vec{q})\exp(i\vec{q}\cdot\vec{r}-iEt),
\label{expand}
\end{eqnarray}
%%%%%
where $\varphi$ and $h$ are defined in Appendix. $\alpha_r, \beta_r$ 
and $\gamma_r$ are some constants. The subscripts $u, d$ and $s$ 
describe flavor and color of quarks in obvious way. 

Plugging (\ref{expand}) into (\ref{eom}), using the bispinor algebraic relations given
in Appendix and assuming constant value of the gap parameter $\Delta$, one obtains the
wavefunction describing the quasiparticle excitations of given energy $E$ in CFL phase:
%%%%%
\begin{eqnarray}
\Psi(t, \vec{r}) \equiv \left(\begin{array}{c} 
\psi^{u}_{red}\\ \psi^{d}_{green}\\ \psi^{s}_{blue}\\ 
\psi^{u\dagger{\rm T}}_{red}\\ \psi^{d\dagger{\rm T}}_{green}\\ \psi^{s\dagger{\rm T}}_{blue}\\ 
\end{array}\right)
=\left[ A\left(\begin{array}{c} 
\frac{E+\xi}{\Delta^*}\varphi^{u}_{\uparrow R}\\ 
-\frac{E+\xi}{\Delta^*}\varphi^{d}_{\uparrow R}\\ 0\\ 
-h^{u\dagger{\rm T}}_{\downarrow L}\\ h^{d\dagger{\rm T}}_{\downarrow L}\\ 0\\ 
\end{array}\right)e^{i\vec{q}_1\cdot\vec{r}}
+ B\left(\begin{array}{c} 
\frac{E-\xi}{\Delta^*}\varphi^{u}_{\uparrow R}\\ 
-\frac{E-\xi}{\Delta^*}\varphi^{d}_{\uparrow R}\\ 0\\ 
-h^{u\dagger{\rm T}}_{\downarrow L}\\ h^{d\dagger{\rm T}}_{\downarrow L}\\ 0\\
 \end{array}\right)e^{i\vec{q}_2\cdot\vec{r}} 
+ C\left(\begin{array}{c} 
\frac{E+\xi}{\Delta^*}\varphi^{u}_{\uparrow R}\\ 0\\
-\frac{E+\xi}{\Delta^*}\varphi^{s}_{\uparrow R}\\  
-h^{u\dagger{\rm T}}_{\downarrow L}\\ 0\\ h^{s\dagger{\rm T}}_{\downarrow L}\\  
\end{array}\right)e^{i\vec{q}_1\cdot\vec{r}} +\right. \nonumber \\
\left. + D\left(\begin{array}{c} 
\frac{E-\xi}{\Delta^*}\varphi^{u}_{\uparrow R}\\ 0\\
-\frac{E-\xi}{\Delta^*}\varphi^{s}_{\uparrow R}\\  
-h^{u\dagger{\rm T}}_{\downarrow L}\\ 0\\ h^{s\dagger{\rm T}}_{\downarrow L}\\  
\end{array}\right)e^{i\vec{q}_2\cdot\vec{r}}
+ F\left(\begin{array}{c} 
\frac{2\Delta}{E-\zeta}\varphi^{u}_{\uparrow R}\\ 
\frac{2\Delta}{E-\zeta}\varphi^{d}_{\uparrow R}\\
\frac{2\Delta}{E-\zeta}\varphi^{s}_{\uparrow R}\\   
h^{u\dagger{\rm T}}_{\downarrow L}\\ 
h^{d\dagger{\rm T}}_{\downarrow L}\\ 
h^{s\dagger{\rm T}}_{\downarrow L}\\  \end{array}\right)e^{i\vec{p}_1\cdot\vec{r}}
+ G\left(\begin{array}{c} 
\frac{2\Delta}{E+\zeta}\varphi^{u}_{\uparrow R}\\ 
\frac{2\Delta}{E+\zeta}\varphi^{d}_{\uparrow R}\\
\frac{2\Delta}{E+\zeta}\varphi^{s}_{\uparrow R}\\   
h^{u\dagger{\rm T}}_{\downarrow L}\\ 
h^{d\dagger{\rm T}}_{\downarrow L}\\ 
h^{s\dagger{\rm T}}_{\downarrow L}\\  \end{array}\right)e^{i\vec{p}_2\cdot\vec{r}} 
\right] \exp{(-iEt)}
\label{solution}
\end{eqnarray}
%%%%%
where $\xi = \sqrt{E^2-\Delta^2}$ and $\zeta = \sqrt{E^2-4\Delta^2}$.
$q_{1,2} = \sqrt{(\mu \pm \xi)^2-m^2}$ and $p_{1,2} = \sqrt{(\mu \pm \zeta)^2-m^2}$.
$A, B, C, D, F$ and $G$ are arbitrary constants. Similar expression is obtained
for the opposite spin content.

Let us set up our physical problem of the quark scattering at the QGP/CFL interface.
We suppose here that there is plane boundary of the interface at $z = 0$. The boundary
is defined as the step function $\Delta(z) = \Delta\Theta (z)$. 
Then we try to find out the solutions of eqs.(\ref{eom}) by
requiring the wavefunctions on both sides (QGP and CFL) to match at $z = 0$.

Let us assume the ``red'' up quark with given energy $E$ falls at the plane boundary from
the left $z<0$ (QGP phase). Then the wavefunction (for $\Delta = 0$) takes the form:
%%%%%%%%%%%%%%%
\begin{equation}
\Psi_{<}(t, z) = \left(\begin{array}{c} 
\varphi^u_{\uparrow R}e^{ik_1 z}+ H\varphi^u_{\uparrow R}e^{-ik_1 z} \\
J\varphi^d_{\uparrow R}e^{-ik_2 z}\\
K\varphi^s_{\uparrow R}e^{-ik_3 z}\\
Lh^{u\dagger{\rm T}}_{\downarrow L}e^{ik_4 z}\\
Nh^{d\dagger{\rm T}}_{\downarrow L}e^{ik_5 z}\\
Ph^{s\dagger{\rm T}}_{\downarrow L}e^{ik_6 z}\end{array}
\right)\exp{(-iEt)},
\label{qgp}
\end{equation}
%%%%%%%%%%%
where $H, J, K, L, N$ and $P$ denote the amplitudes of the reflection of the particle and holes
and $k_1 = k_2 = k_3 \equiv k = \sqrt{(\mu + E)^2-m^2}$ 
and $k_4 = k_5 = k_6 \equiv l = \sqrt{(\mu - E)^2-m^2}$, respectively. For $z >0$,
the quasiparticle excitations in CFL phase are described by the wavefunction $\Psi_{>}(t, z)$
given by the expression (\ref{solution}). The continuity conditions to match the wavefunctions
at the interface are of the form:
%%%%%%%%%%%%%%%
\begin{equation}
\Psi_{<}(t, z = 0) = \Psi_{>}(t, z = 0). 
\label{bc}
\end{equation}
%%%%%%%%%%%
Using this condition one can find the amplitude of the scattering process at
leading order in $\mu$ expansion in the massless limit:
%%%%%
\begin{eqnarray}
A &=& C = \frac{\Delta}{3(E+\xi)}+ {\cal O}\left(\frac{1}{\mu}\right), \nonumber \\
F &=& \frac{2\Delta}{3(E+\zeta)}+ {\cal O}\left(\frac{1}{\mu}\right), \nonumber \\
L &=& \frac{E-\zeta}{6\Delta}-\frac{2\Delta}{3(E+\xi)} + {\cal O}\left(\frac{1}{\mu}\right), \nonumber \\
N &=& P =\frac{E-\zeta}{6\Delta}+\frac{\Delta}{3(E+\xi)} + {\cal O}\left(\frac{1}{\mu}\right)
\label{amplitude}
\end{eqnarray}
%%%%%
and other coefficients vanish in the limit where $\Delta, E << \mu$. It is worth mentioning more
about the result obtained here. By the scattering of the red up quark at the interface, holes
of the green down quark and the blue strange quark can be reflected into the QGP phase and 
the quasiparticles with momenta $q_1$ and $p_1$ can propagate in the CFL phase. This is
the similar property which has been observed in the QGP/2SC interface\cite{mariusz1}. However,
unlike the QGP/2SC case, hole of the red up quark is also reflected in the QGP phase.
This can be interpreted as follows. When the red up quark falls toward the boundary ($z = 0$),
it takes another quark together in order to make a Cooper pair.
In the 2SC case, the Cooper pair takes the form $<ud>$ so 
the up quark takes only the down quark leaving the d-hole in QGP phase. On the other hand,
in the CFL phase, we have two kinds of gaps. One is the gap for octet excitation and
the other is for singlet excitation. In the former case, the condensate takes the form
$<ud>$, $<us>$ and $<ds>$ similar to the 2SC case and $<uu-dd-2ss>$. In the latter case,
the condensate takes the form $<uu+dd+ss>$. This means that the
up quark has take not only $d$ and $s$, but also $u$ quarks to make a pair.
In this way we are left with these holes of $u, d$ and $s$-type.

Finally let us calculate the probability current. 
The conserved probability current is given by
$\vec{j} = \Psi^{\dagger}_{<}\vec{\alpha}\Psi_{<} = \Psi^{\dagger}_{>}\vec{\alpha}\Psi_{>}$.
Using eqs.(\ref{solution}) (or (\ref{qgp}))  and (\ref{amplitude}) one finds:
%%%%%%%%
\begin{eqnarray}
j_z = \left\{ \begin{array}{lcr} 0 &  {\rm for}  & E <|\Delta|,\\
2\mu\frac{4}{3}\frac{\xi}{E+\xi} &  {\rm for} & |\Delta|< E < 2|\Delta|,\\
2\mu\frac{4}{3}\frac{\xi}{E+\xi}+2\mu\frac{2}{3}\frac{\zeta}{E+\zeta} &
{\rm for}  & E > 2|\Delta|. \end{array} \right.
\label{current}
\end{eqnarray}
%%%%%%%%
The result (\ref{current}) is interpreted as follows: If the incoming up quark has energy below
$\Delta$ it cannot excite quasiparticles in CFL phase. However if the up quark has energy
between $|\Delta|$ and $2|\Delta|$, it excites quasiparticles which are separated from the
vacuum by the gap $|\Delta|$. If the up quark possesses energy above $2|\Delta|$ it excites
quasiparticles with the gap $2|\Delta|$ as well as those with $|\Delta|$. This result might
be essential when we consider physics of the Neutron Star.

%%%%%%%%%%%%%%%%%%%%%%%%%%%%%%%%%%%%%%%%%%%%%%%%%%%%%%%%%%%%%%%%%%%%%%%%%%%%%%%%%%%%%%%%%%%%
\section{Conclusions}
In this paper we considered the Andreev reflection of quarks
from the QGP/CFL interface.
There are two different types of reflection. One is similar
to the Andreev reflection from the 2SC color superconductor where one
hole (of different color and flavor than incoming particle) is reflected
toward the QGP phase.  However in the CFL phase there is also
the possibility of the reflection of three holes.
This is connected to the fact that in CFL phase
there exist two independent Cooper pair structures:
one related to octet and one related to singlet representations of 
$SU(3)$ group. In the basis chosen in our paper
the example of such a process is the reflection of holes of
$u$ red quark, $d$ green quark and $s$ blue quark from the
incoming particle of $u$ red type. 

The importance of the Andreev reflection phenomenon for the
transport processes is given by
the equation (\ref{current}). From this equation it is seen
that the probability current is strongly suppressed
by the Boltzman factor for energy of 
incoming quarks which are usually
of the order of temperature inside the Nuclear Stars
(and $T<<|\Delta |$). This process is more important
for the QGP/CFL interface than for the QGP/2SC
because in the case of CFL phase all quarks 
in all colors are paired. The existence of
massless Nambu-Goldstone mode connected to the
symmetry breaking, as was shown, does not influence
the transport processes through the
interface in the leading order.

The interior of the Neutron Star is a complicated and not
completely well-known object and in particular can contain
many (or single) interfaces. In that situation the
dynamics of the matter propagation would be
strongly affected by the process of Andreev reflection.
But this subject remains to be done.\\
 
{\bf Acknowledgement} We would like to thank Krishna Rajagopal and Sanjay Reddy
for interesting discussions.\\
M. S. was supported by a fellowship from the Foundation for Polish Science.
M. S. was also supported in part by Polish State Committee for Scientific Research, 
grant no. 2P 03B 094 19. M. T. was supported in part by Grant-in-Aid for Scientific
Research from Ministry of Education, Science, Sports and Culture of Japan (No. 3666).
%%%%%%%%%%%%%%%%%%%%%%%%%%%%%%%%%%
\begin{center}
{\large{\bf Appendix}}
\end{center}

Let us defined the bispinors $\varphi_{r,R}(\vec{k})$ and
$\varphi_{r,L}(\vec{k})$ through the equations:
\begin{eqnarray}
(\alpha\cdot\vec{k}+m\gamma_0-\mu )\varphi_{r,R}(\vec{k}) = 
\epsilon\varphi_{r,R}(\vec{k})\\\nonumber
\varphi^{\dagger }_{r,L}(\vec{k})(\alpha\cdot\vec{k}+m\gamma_0-\mu )= 
\epsilon\varphi^{\dagger}_{r,L}(\vec{k})
\end{eqnarray}
where $r$ describes the spin and momentum $\vec{k}$ can be in
general complex vector. From this reason one has to distinguish
between the right- and left-handed eigenvectors, which is
denoted by the capital letters $L,R$. For $\vec{k}$ complex
$\varphi^{\dagger }_{r,L}$ is not hermitian conjugate to $\varphi_{r,R}$.
The solution of the equations takes the form:
\begin{eqnarray}
\varphi_{r,R}(\vec{k})=\left( 
\begin{array}{c}
\sqrt{m+\epsilon +\mu} \chi_r \\
\frac{\vec{\sigma }\cdot\vec{k}}{\sqrt{m+\epsilon +\mu }}\chi_r \\
\end{array}
\right) \\\nonumber
\varphi^\dagger_{r,L}(\vec{k})=\left(\bar{\chi}_{r}^\dagger\sqrt{m + \epsilon + \mu},
\bar{\chi}_{r}^\dagger\frac{\vec{\sigma }\cdot\vec{k}}{\sqrt{m+\epsilon +\mu }}\right)
\end{eqnarray}
where $\chi_r ,\bar{\chi}_{r}^\dagger $ are spinors and where 
$\epsilon = \sqrt{\vec{k}^2+m^2}-\mu $\footnote{There is another
solution with $\epsilon = -\sqrt{\vec{k}^2+m^2}-\mu $ which is not interesting
for our purposes}.
The spinors can be defined in the helicity basis:
\begin{eqnarray}
\vec{\sigma }\cdot\vec{k}\chi_{\uparrow ,\downarrow }= 
\pm k\chi_{\uparrow ,\downarrow }\\\nonumber
\bar{\chi}_{\uparrow ,\downarrow }^\dagger\vec{\sigma }\cdot\vec{k}
= \pm k\bar{\chi}_{\uparrow ,\downarrow }^\dagger
\end{eqnarray} 
where $k=\sqrt{\vec{k}^2}$.
Let us also define the additional bispinors:
\begin{eqnarray}
h^\dagger_{r,L}(\vec{k})(\alpha\cdot\vec{k}- m\gamma_0 + \mu ) = 
\bar{\epsilon } h^\dagger_{r,L}(-\vec{k})\\\nonumber
(\alpha\cdot\vec{k}- m\gamma_0 + \mu )h_{r,R}(-\vec{k}) = \bar{\epsilon }h_{r,R}(-\vec{k})
\end{eqnarray}
where $\bar{\epsilon }=-\epsilon $ and bispinors are given by formulae: 
\begin{eqnarray}
h_{r,R}(-\vec{k})=\left( 
\begin{array}{c}
\sqrt{m-\epsilon +\mu} \chi_r \\
\frac{\vec{\sigma }\cdot\vec{k}}{\sqrt{m-\epsilon +\mu }}\chi_r \\
\end{array}
\right) \\\nonumber
h^\dagger_{r,L}(-\vec{k})=\left({\bar{\chi}_r}^\dagger\sqrt{m - \epsilon + \mu},
-{\bar{\chi}_r}^\dagger\frac{\vec{\sigma }\cdot\vec{k}}{\sqrt{m-\epsilon +\mu }}\right)
\end{eqnarray}
The above defined bispinors fulfill simple algebraic relations
which are useful in the calculations:
\begin{eqnarray}
\varphi_{r,L}^\dagger\varphi_{s,R}=h_{r,L}^\dagger h_{s,R}=2\sqrt{k^2+m^2}\delta_{rs}\\\nonumber
\varphi_{s,L}^\dagger C\gamma_5 h_{r,L}^{\dagger\,T} h_{s,R} =
\varphi_{s,R}^{T} C\gamma_5 h_{r,R} = 2\sqrt{k^2+m^2}
\left\{
\begin{array}{lr}
-1 & s=\uparrow\,r=\downarrow \\
1 & s=\downarrow\, r=\uparrow 
\end{array}
\right.
\end{eqnarray}

The bispinors defined above has simple physical meaning in the QGP phase.
The wavefunction:
\begin{equation}
\psi (t,\vec{r}) = \varphi_{\uparrow\, R}(\vec{k})\exp{(-i\epsilon t+i\vec{k}\cdot\vec{r})}
\end{equation}
describes the particle of spin projection up, velocity $\vec{v} = \frac{\vec{k}}{E}$, 
where $E = \sqrt{{\vec{k}}^2+m^2}$ and energy $\epsilon $ above
the Fermi Sea. From the other hand the wavefunction:
\begin{equation}
\psi^\dagger (t,\vec{r}) = 
h^{\dagger}_{\downarrow\, L}(-\vec{k})\exp{(-i\epsilon t+i\vec{k}\cdot\vec{r})}
\end{equation}
describes the hole of spin projection down, velocity $-\vec{v}$, 
and energy $\epsilon$ below the Fermi Sea.

\end{document}